\documentclass[aip,pre,preprint,superscriptaddress,showpacs]{revtex4}

\usepackage{amssymb}
\usepackage{color}
\usepackage{graphicx}
\usepackage{amsmath}
\usepackage{times}
\usepackage{bm}
\usepackage{float}

\begin{document}

\title{Synchronization of coupled metronomes on two layers}

\author{Jing Zhang}
\affiliation{School of Physics and Information Technology, Shaanxi Normal University, Xi'an 710062, China}

\author{Yizhen Yu}
\email[Email address: ]{yzyu@snnu.edu.cn}
\affiliation{School of Physics and Information Technology, Shaanxi Normal University, Xi'an 710062, China}

\author{Xingang Wang}
\email[Email address: ]{wangxg@snnu.edu.cn}
\affiliation{School of Physics and Information Technology, Shaanxi Normal University, Xi'an 710062, China}

\begin{abstract}
Coupled metronomes serve as a paradigmatic model for exploring the collective behaviors of complex dynamical systems, as well as a classical setup for classroom demonstrations of synchronization phenomena. Whereas previous studies of metronome synchronization have been concentrating on symmetric coupling schemes, here we consider the asymmetric case by adopting the scheme of layered metronomes. Specifically, we place two metronomes on each layer, and couple two layers by placing one on top of the other. By varying the initial conditions of the metronomes and adjusting the friction between the two layers, a variety of synchronous patterns are observed in experiment, including the splay synchronization (SS) state, the generalized splay synchronization (GSS) state , the anti-phase synchronization (APS) state, the in-phase delay synchronization (IPDS) state, and the in-phase synchronization (IPS) state. In particular, the IPDS state, in which the metronomes on each layer are synchronized in phase but are of a constant phase delay to metronomes on the other layer, is observed for the first time. In addition, a new technique based on audio signals is proposed for pattern detection, which is more convenient and easier to apply than the existing acquisition techniques. Furthermore, a theoretical model is developed to explain the experimental observations, and is employed to explore the dynamical properties of the patterns, including the basin distributions and the pattern transitions. Our study sheds new lights on the collective behaviors of coupled metronomes, and the developed setup can be used in the classroom for demonstration purposes.
\end{abstract}
\date{\today}
\maketitle

\section{Introduction}

Synchronization behavior of coupled oscillators is not only a fascinating manifestation of self-organization that nature uses to orchestrate essential processes of life, such as the beating of the heart, but also common in man-made systems, including Josephson junction, semiconductor laser, etc \cite{Kuramoto,PRK:2001,Strogatz:2003}. The first scientific record of synchronization can be traced to the letter from Christian Huygens to Moray in 1665, where the anti-phase synchronization (APS) of two pendulum clocks hung from a common beam supported by two chairs has been described \cite{Huygens}. In the past decades, with the blooming of nonlinear science and complex systems, synchronization behaviors in coupled oscillators have been extensively studied in a variety of fields, and are regarded as being closely connected to the functionality and operation of many realistic systems \cite{2012 Kapitaniak,NETSYN:REV0,NETSYN:REV1,NETSYN:REV2,LZH:FoP,GSG:FoP}. For two coupled oscillators, the research interest is mainly focusing on the entrainment of the oscillator trajectories, in which a variety of synchronization forms have been revealed, including complete synchronization, in-phase synchronization (IPS), APS, generalized synchronization, delay synchronization, etc \cite{PRK:2001}. While for an ensemble of coupled oscillators, the research interest has been mainly focusing on the formation of various synchronous patterns, and also the transition among the patterns \cite{NETSYN:REV0}. Recently, stimulated by the discovery of the small-world and scale-free features in many natural and man-made complex networks, the synchronization behaviors of complex networks have received considerable attention \cite{NETSYN:REV1,NETSYN:REV2}. 

The experiment of  two coupled pendulums studied by Huygens has been revisited actively in recent years, for its easy realization in experiment and the rich dynamical phenomena it offers \cite{2012 Kapitaniak}. In 2002, Bennett \emph{et al}. repeated the experiment by a redesigned device where two pendulum clocks are hang on a heavy supporter mounted on a low-friction wheeled cart. It is found that the APS state discovered by Huygens occurs only when the natural frequencies of the clocks are of small mismatch \cite{2002 bennett}. Pantaleone simplified the experiment by placing two metronomes on a freely moving board supported by two empty cans \cite{2002 Pantaleone}. Meanwhile, a theoretical model has been developed to describe the motions of the metronomes, which is able to reproduce both the IPS and APS states. Wu \emph{et al.} studied the basin distribution of IPS and APS states, and found that the attracting basin of the APS state can be efficiently enlarged by increasing the rolling friction \cite{2012 Wu,2014 Wu}. Song \emph{et al.} studied experimentally the phase synchronization of two coupled nonidentical metronomes, and found the frequencies of the metronomes can be looked onto an irrational ratio \cite{2015 Song}.
 
When three pendulums (metronomes) are coupled, an intriguing phenomenon is the formation of synchronous patterns. By coupling three identical pendulums, Czolczynski \emph{et al.} observed the splay synchronization (SS) state where the phase difference between any two pendulums is fixed at $2\pi/3$. By a CCD acquisition system, Hu \emph{et al.} confirmed this SS state in experiment \cite{2013 Hu}. By the same acquisition system, Jia \emph{et al.} successfully generated in experiment the triplet synchronization state predicted theoretically by Kralemann \emph{et al} \cite{TripletSyn2013, 2015 Jia}. When more than three metronomes are coupled, a general phenomenon is that the metronomes can be self-organized into different synchronous clusters. In cluster synchronization, metronomes within the same cluster are synchronized, but are not if they belong to different clusters.  By hanging pendulum clocks from an elastically fixed horizontal beam, Czolczynski \emph{et al.} studied the synchronization behavior of up to $11$ pendulums, and found that the synchronous clusters are dependent of the network size and satisfy certain symmetries \cite{2009 Czolczynski}. The synchronization behavior of an ensemble of pendulums (metronomes) has been also investigated. Ulrichs \emph{el al.} studied numerically up to $100$ coupled metronomes, and found that the onset of synchronization is similar to that of Kuramoto model \cite{2009 Ulrichs}. Martens \emph{et al.} studied the synchronization of a hierarchically coupled metronome network, and observed the chimera-like state \cite{2013 Martens}. Boda \emph{et al.} analyzed the onset of synchronization experimentally by placing metronomes on a freely rotating disc, and found the Kuramoto-type phase transition too \cite{2013 Boda}. Recently, coupling metronomes with a ring structure,  Kapitaniak \emph{et al.} observed the imperfect chimera state \cite{2014 Kapitaniak}. 

The synchronization behaviors of coupled oscillators are crucially dependent on the symmetry of the underlying coupling structure \cite{PRK:2001}. For systems of symmetric coupling, the synchronous patterns are generally generated by the mechanism of symmetry breaking \cite{PartialSyn:Hasler,PartialSyn:ZY,Chimera:2004}. While symmetry breaking explains the patterns arisen in regular medium, it fails to describe the formation of patterns in complex systems, e.g., the synchronous patterns on complex networks \cite{SymmNet:2013,SymmNet:2014,SymmNet:Pecora,SymmNet:Motter}. For the sake of simplicity, in previous studies of metronome synchronization the metronomes are normally placed on the same board (or hung to the same beam for coupled pendulums). In such a coupling scheme, the system possesses a high-order symmetry (global-coupling structure), as the system dynamics is kept unchanged by exchanging the metronomes. Considering the significant influence of coupling symmetry on the formation of patterns \cite{SymmNet:2013,SymmNet:2014,SymmNet:Pecora,SymmNet:Motter}, a natural question therefore is whether synchronous patterns can be observed in asymmetrically coupled metronomes. The answer to this question is of both theoretical and practical significance, as perfect symmetries are impossible in realistic situations. In the present study, we move forward a step from the traditional studies of symmetrically coupled metronomes by deteriorating the coupling symmetry, and explore the new synchronization behaviors arisen in asymmetrically coupling metronomes. 

The rest of the paper is organized as follows. In Sec. II, we will introduce the experimental setup, and also the new technique used for detecting the synchronous patterns. In Sec. III, we will present the experimental results. In Sec. IV, we will construct the theoretical model, and, by numerical simulations, explore the dynamical properties of the synchronous patterns, including their basin distributions and the pattern transitions. Discussions and conclusion will be given in Sec. V.

\section{Experimental setup and measure technique}

\begin{figure}[tpb]
\includegraphics[width=\columnwidth]{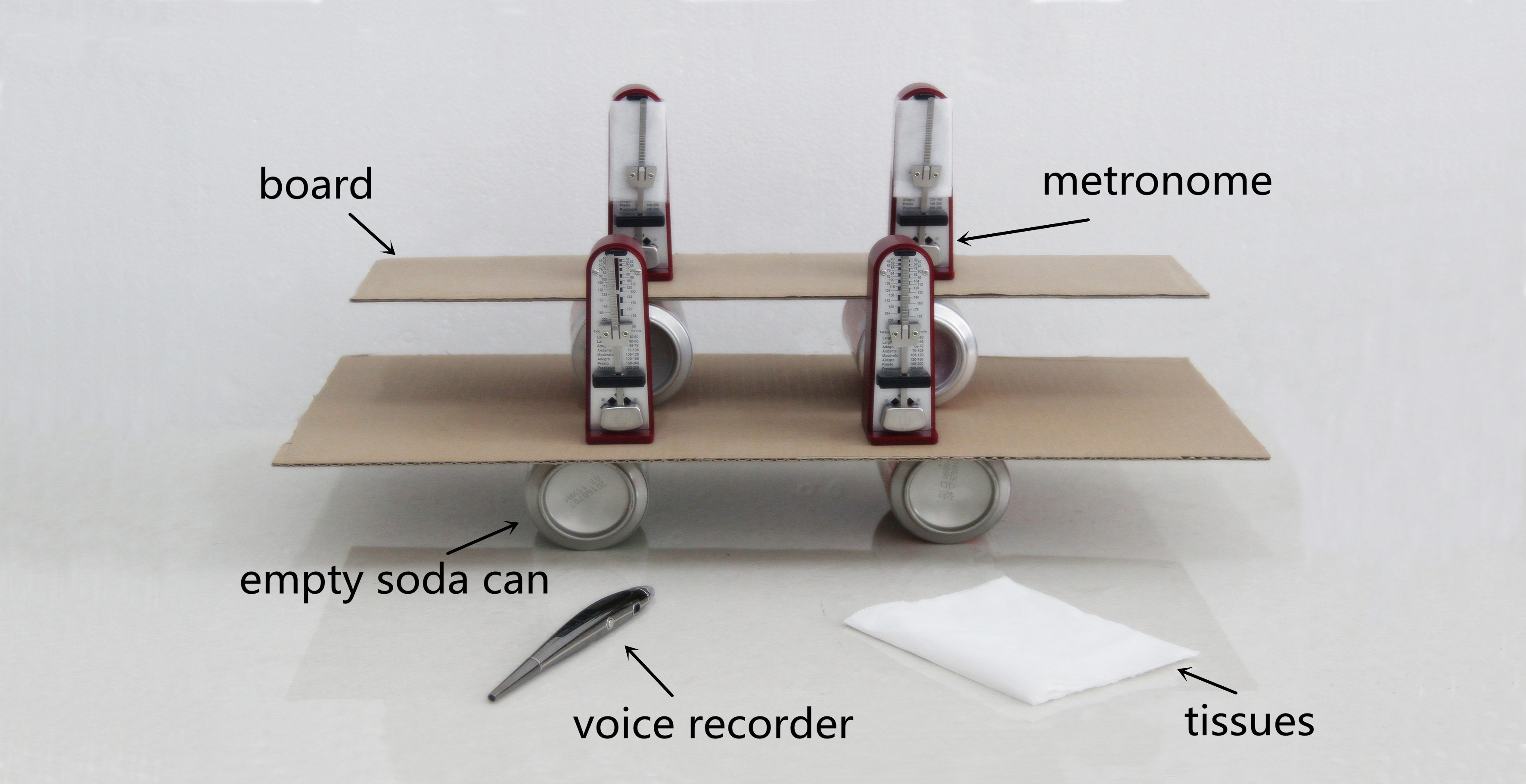}
\caption {The experimental setup.}
\label{fig1}
\end{figure}

Our experimental setup of layered metronomes is shown in Fig. \ref{fig1}. Following the design of Pantaleone \cite{2002 Pantaleone}, we construct each layer by placing two identical metronomes on a light cardboard, and support the cardboard by two empty soda cans. The size of the bottom cardboard is larger than the top one, so as to provide sufficient space for a free movement of the top layer. The whole setup is then placed on the desk. All four metronomes (Wittner's Taktell, Series 890) are of nearly identical parameters (parameter mismatch less than 1\%), with the energy supplied by a hand-wound spring. The frequency of the metronome can be adjusted by changing the position of the mass on the pendulum bob, ranging from 40 ticks per minute (largo) to 208 ticks per minute (prestissimo). This experimental setup of coupled metronomes is easy to assemble and is able to generate robust synchronous patterns within a few of seconds, making it an excellent experiment for classroom demonstration.

Two special techniques are used in our experimental studies. First, following Ref. \cite{2012 Wu}, we pad some paper tissues below the soda cans, so as to modify the frictions between the two layers, as well as the friction between the bottom layer and the desk. Second, a voice recorder (HYUNDAI, HYV-B600) is used to collect the audio signals of the metronomes. To differentiate the audio signals released from metronomes at different layers, we cover a small pice of cotton cloth on the front face (the side with the speaker) of the metronomes on the upper layer. Meanwhile, to keep the mass of all the metronomes identical, metronomes on the lower layer are also attached with the same-size cloth at the reverse side.  As such, the ``tick" sounds of the upper-layer metronomes have the reduced intensity in the audio signals. We download the recorded audio signals from the recorder to the computer, from which the synchronous patterns of the metronomes will be analyzed. In the meantime, a smart cell phone with camera function is used to record the video information, which will be used as the auxiliary method to identify further the synchronous patterns. (Two different patterns may have the same audio signals, as each metronome ``ticks" twice in a cycle.)

The detection of synchronous patterns in coupled metronomes has long been a challenging problem in experimental studies \cite{2012 Kapitaniak,2013 Martens,2014 Kapitaniak}. In Ref. \cite{2013 Hu}, the authors developed a sophisticated CCD acquisition system to detect the SS state, which is able to measure the synchronization relationship among the metronomes with a high precision \cite{2015 Jia}. Similar video-based techniques have been used in detecting the collective behavior of an ensemble of coupled metronomes \cite{2013 Martens,2014 Kapitaniak}. These techniques, however, rely heavily on the camera functions and use some modern techniques of signal analysis, making it difficult to be applied in classroom education. Here we propose to detect the synchronous patterns from the audio signals, which, comparing to the previous video-based methods, is more convenient and efficient. As the metronomes have identical natural frequency and each metronome ticks twice in a cycle (at the highest points), we thus can infer from the audio signals the synchronous behaviors by just listening the ``rhythm". Moreover, as the speakers of the upper metronomes are covered by cotton cloths, the signals released from the upper metronomes can be identified easily. As such, we can also detect the detailed configuration of the synchronous patterns, i.e., which pair of metronomes in the experiment are synchronized. 

\section{Experimental results}

\begin{figure}
\includegraphics[width=\columnwidth]{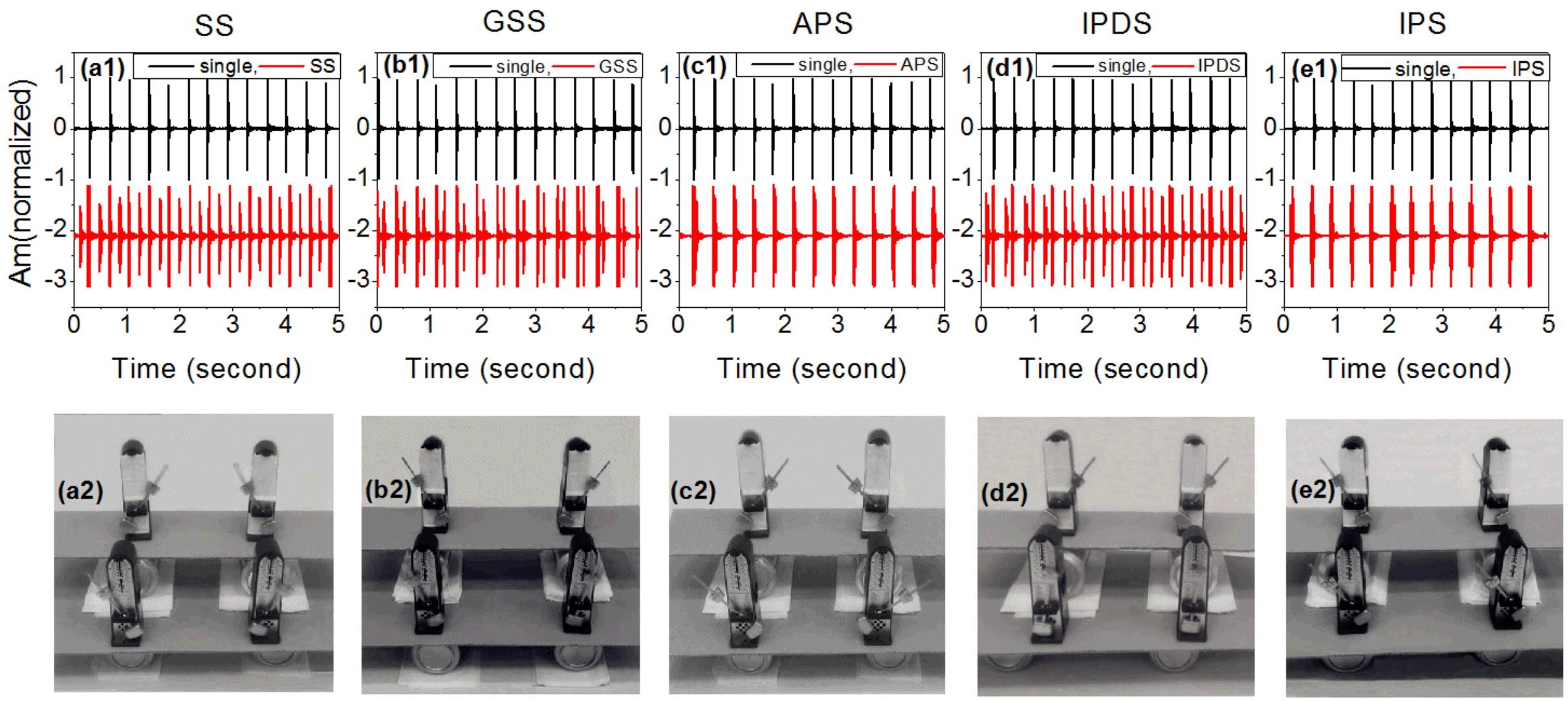}
\center
\caption {(Color online) Synchronous patterns observed in experiment. (a1)$\sim$(e1) The recorded audio signals. The upper panel in each subplot (black colored) is taken from an isolated metronome, which is used as the reference signal for pattern analysis. (a2)$\sim$(e2) Snapshots of the experimental setup. (a1,a2) The splay synchronization (SS) state. (b1,b2) The generalized splay synchronization (GSS) state. (c1,c2) The anti-phase synchronization (APS) state. (d1,d2) The in-phase delay synchronization (IPDS) state. (e1,e2) The in-phase synchronization (IPS) state. See context for more details about each pattern.}
\label{fig2}
\end{figure}

Setting the natural frequency of the (four) metronomes as ``allegro" (which corresponds to 160 ticks per minute), we study experimentally the formation of different synchronous patterns by the setup presented in Fig. \ref{fig1}. To start, we pad $4$ layers of paper tissues below each can, pull the pendulums away from their equilibria by a small angle, and then release them without giving it any initial momentum. After a transient period of $T=50$ seconds, we record the audio and video signals for $5$ seconds. We then download the audio and video signals to the computer, and analyze synchronous patterns. This process is repeated for $100$ runs, and $3$ different synchronous patterns are observed in total, which are plotted in Fig. \ref{fig2}(a-c).  

Fig. \ref{fig2}(a) shows the SS state. The audio signals of the metronomes for this state are plotted in Fig. \ref{fig2}(a1) (the lower panel). As the reference, we plot in Fig. \ref{fig2}(a1) also the audio signal taken from an isolated metronome (the upper panel). It is seen that the phases of the metronomes are equally separated in the space, i.e., by a constant phase difference of $\Delta \phi=\pi/2$, showing the feature of SS \cite{2013 Hu}. In addition, the audio signals of the upper metronomes, which have the reduced intensity, can be straightforwardly distinguished from that of the lower metronomes. For the reduced audio intensity of the upper metronomes, more information about the synchronous patterns can be inferred. Specifically, comparing with the reference signal of isolated metronome (the upper panel), it is seen that the weak (from the upper metronomes) and strong (from the lower metronomes) signals are shown up alternatively with the fixed phase difference $\Delta \phi=\pi/2$. We thus infer from Fig. \ref{fig2}(a) that metronomes on each layer reach anti-phase synchronization (out of phase by $\pi$), and the phase difference between the upper and lower metronomes is fixed to $\pi/2$. By checking the corresponding video [Fig. \ref{fig2}(a2)], it is shown that, indeed, metronomes on each layer reach the anti-phase synchronization, and the phases of the upper and lower metronomes are differentiated by about $\pi/2$.   

Fig. \ref{fig2}(b) shows the generalized splay synchronization (GSS) state, which is generated by changing the initial phases of the metronomes. In GSS, metronomes on the same layer reach anti-phase synchronization, but, different from the SS state, the phase difference between the upper (weak audio signals) and lower metronomes (strong audio signals) is locked to some value between $0$ and $\pi/2$, showing the feature of GSS \cite{2001Zhan}. Fig. \ref{fig2}(c) shows the APS state, generated by another set of random initial conditions. In APS, metronomes on each layer reach anti-phase synchronization, while metronomes on different layers may reach either in-phase or anti-phase synchronization. It is worth mentioning that in SS, GSS and APS, anti-phase synchronization is always established between metronomes on the same layer. The difference between the three states lies in the phase delay between the upper and lower metronomes: $\Delta \phi=\pi/2$ for SS, $\Delta \phi=\pi$ for APS, while $\Delta \phi\in (0,\pi)$ for GSS (with the exact value varying with the initial conditions). In experimental studies, it is noticed that the three states are generated with different probabilities. Comparing to the SS and APS states, the GSS state is more easily generated. 

Removing the paper tissues between the cans and the desk, we observe the in-phase delay synchronization (IPDS) state [Fig. \ref{fig2}(d)]. From only the audio signals, we can not tell the difference between the IPDS [Fig. \ref{fig2}(d1)] and GSS [Fig. \ref{fig2}(b1] states, as in both states the upper metronomes are delayed by a constant phase to the lower metronomes. Yet, by checking the video signal, it is found that in IPDS the metronomes within each layers reach in-phase synchronization, instead of anti-phase synchronization as for GSS. Padding more paper tissues between the cans and the lower board ($10$ pieces of paper tissues), the IPS state is observed [Fig. \ref{fig2}(e)], in which all metronomes are oscillating in step. The audio signal of IPS is identical to APS, yet the video signal reveals again the difference [Fig. \ref{fig2}(e2)]. We wish to note that with the same frictions used in generating SS (GSS and APS), the IPDS and IPS states is not observed in experiment by changing the initial conditions (for $100$ tries). To generate IPDS (IPS), we need to adjust the frictions between the cans and the desk (bottom board). This observation suggests that the stability of the patterns is dependent on the frictions.

\section{Theoretical model and numerical results}

\begin{figure}
\includegraphics[width=\columnwidth]{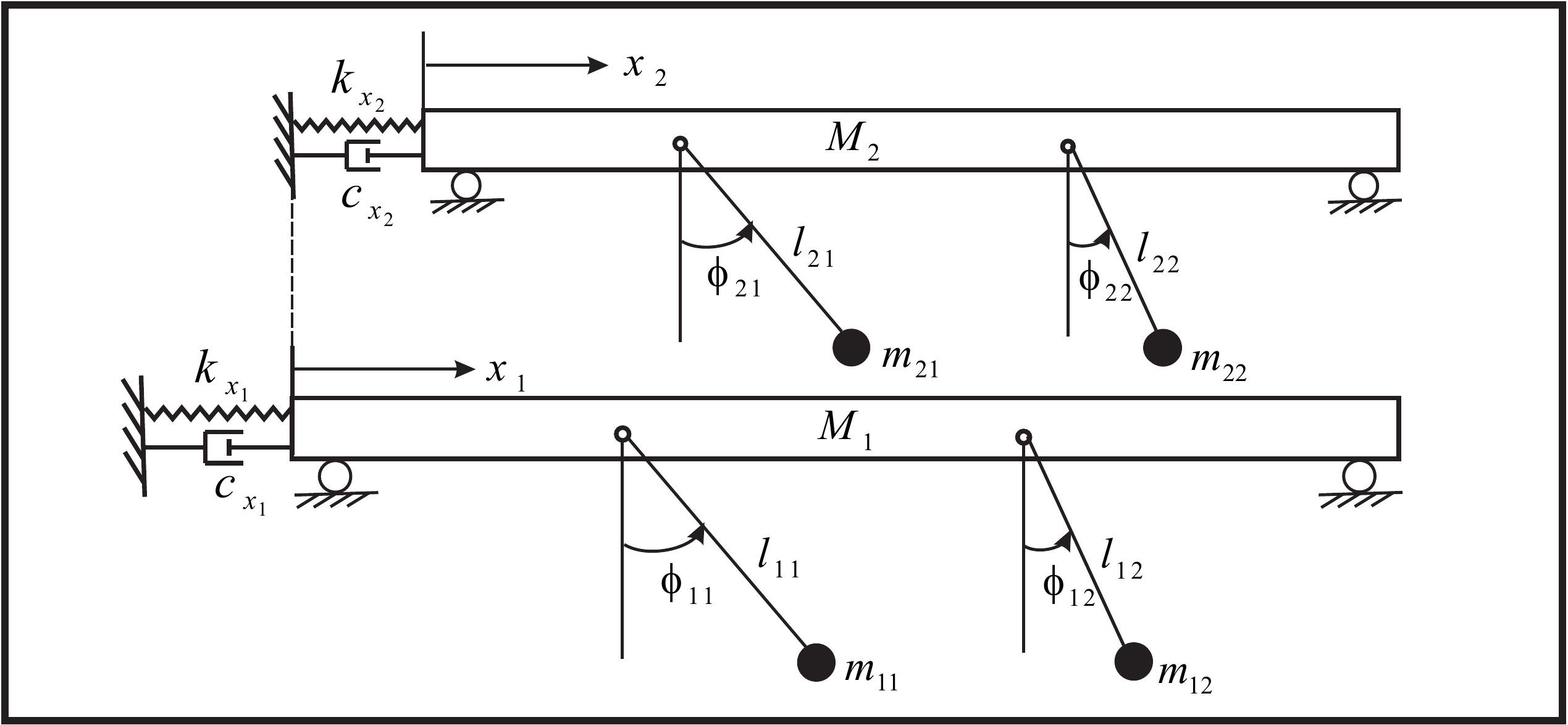}
\caption {Theoretical model of the experimental setup.}
\label{fig3}
\end{figure}

We next investigate the synchronization behavior of layered metronomes theoretically. Following Ref. \cite{2009 Czolczynski}, the experimental setup shown in Fig. \ref{fig1} can be modeled by the sketch plotted in Fig. \ref{fig3}. As shown in Fig. \ref{fig3}, the theoretical model consists of two rigid beams, which correspond to the two cardboards in the experimental setup, and each beam is connected to a linear spring (so as to keep the beams staying around their equilibrium points). The metronomes at each layer are modeled by two point pendulums suspended to the beams. $M_{1}$ and $M_2$ denote, respectively, the mass of the bottom and top cardboards (beams). The displacements of the beams from their corresponding equilibrium points are represented by $x_{1}$ (the bottom beam) and $x_{2}$ (the top beam). The stiffness coefficient of the spring for the bottom (top) beam is $k_{x_{1}}$ ($k_{x_{2}}$), and the friction coefficient is $c_{x_{1}}$ ($c_{x_{2}}$). $m_{ij}$ denotes the mass of the $j$th pendulum at layer $i$, and $\phi_{ij}\in(-\pi, \pi)$ represents the phase of this pendulum. To mimic the energy input of the mechanical metronome, we drive the pendulum by the momentum $D_{ij}$ when $|\phi_{ij}|$ is smaller to a small angle $\gamma_{N}$. Specifically,  $D_{ij}=D$ when $0<\phi_{ij}<\gamma_{N}$ and $\dot{\phi}_{ij}>0$, $D_{ij}=-D$ when $-\gamma_{N}<\phi_{ij}<0$ and $\dot{\phi}_{ij}<0$, otherwise $D_{ij}=0$.

The kinetic energy $E$ and the potential energy $V$ of the system can be written as
\begin{eqnarray}\label{e1}
\begin{split}
E&=\frac{1}{2}M_{1}\dot{x}_{1}^{2}+\frac{1}{2}M_{2}\dot{x}_{2}^{2}+\sum_{j=1}^{2}(\frac{1}{2}m_{1j}(\dot{x}_{1j}^{2}+\dot{y}_{1j}^{2})+\frac{1}{2}m_{2j}(\dot{x}_{2j}^{2}+\dot{y}_{2j}^{2})), \\ 
  V&= -\sum_{j=1}^{2}(m_{1j}gl_{1j}\cos\phi_{1j}+m_{2j}gl_{2j}\cos\phi_{2j}),
\end{split}
\end{eqnarray}
with $g$ the gravity constant and $l_{ij}$ the length of the pendulums. From Eq. (\ref{e1}) we have the Lagrange equations
\begin{eqnarray}\label{e2}
\begin{split}
\frac{d}{dt}(\frac{\partial E}{\partial \dot{x}_{1}})-\frac{\partial E}{\partial x_{1}}+\frac{\partial V}{\partial x_{1}} &=  c_{x_{2}}(\dot{x}_{2}-\dot{x}_{1})-c_{x_{1}}\dot{x}_{1}+k_{x_{2}}(x_{2}-x_{1})-k_{x_{1}}x_{1},\\
 \frac{d}{dt}(\frac{\partial E}{\partial \dot{x}_{2}})-\frac{\partial E}{\partial x_{2}}+\frac{\partial V}{\partial x_{2}} &=  -c_{x_{2}}(\dot{x}_{2}-\dot{x}_{1})-k_{x_{2}}(x_{2}-x_{1}),\\
 \frac{d}{dt}(\frac{\partial E}{\partial \dot{\phi}_{ij}})-\frac{\partial E}{\partial \phi_{ij}}+\frac{\partial V}{\partial \phi_{ij}} &=  -c_{\phi_{ij}}\dot{\phi}_{ij}+D_{ij} \quad  (i,j=1,2),
\end{split}
\end{eqnarray}
with $c_{\phi_{ij}}$ the friction coefficient. From Eqs. (\ref{e1}) and (\ref{e2}), we obtain the following motion equations
\begin{eqnarray}\label{e3}
\begin{split}
(M_{1}+\sum_{j=1}^{2}m_{1j})\ddot{x}_{1}+\sum_{j=1}^{2}m_{1j}l_{1j}(\ddot{\phi}_{1j}\cos\phi_{1j} -\dot{\phi}_{1j}^{2}\sin\phi_{1j})&=c_{x_{2}}(\dot{x}_{2}-\dot{x}_{1})-c_{x_{1}}\dot{x}_{1}
 \\  &+k_{x_{2}}(x_{2}-x_{1})-k_{x_{1}}x_{1}, \\
(M_{2}+\sum_{j=1}^{2}m_{2j})\ddot{x}_{2}+\sum_{j=1}^{2}m_{2j}l_{2j}(\ddot{\phi}_{2j}\cos\phi_{2j}-\dot{\phi}_{2j}^{2}\sin\phi_{2j})
   &= c_{x_{2}}(\dot{x}_{1}-\dot{x}_{2})+k_{x_{2}}(x_{1}-x_{2}), \\
  m_{ij}l_{ij}^{2}\ddot{\phi}_{ij}+m_{ij}l_{ij}\ddot{x}_{i}\cos\phi_{ij}+c_{\phi_{ij}}\dot{\phi}_{ij}+m_{ij}gl_{ij}\sin\phi_{ij}&=D_{ij}, \quad  (i,j=1,2).
\end{split}
\end{eqnarray}
Eqs. (\ref{e3}) describes the dynamics of the system presented in Fig. \ref{fig3}, and will be analyzed numerically in the next section to explore the synchronization behaviors of layered metronomes.

\begin{figure}
\includegraphics[width=\columnwidth]{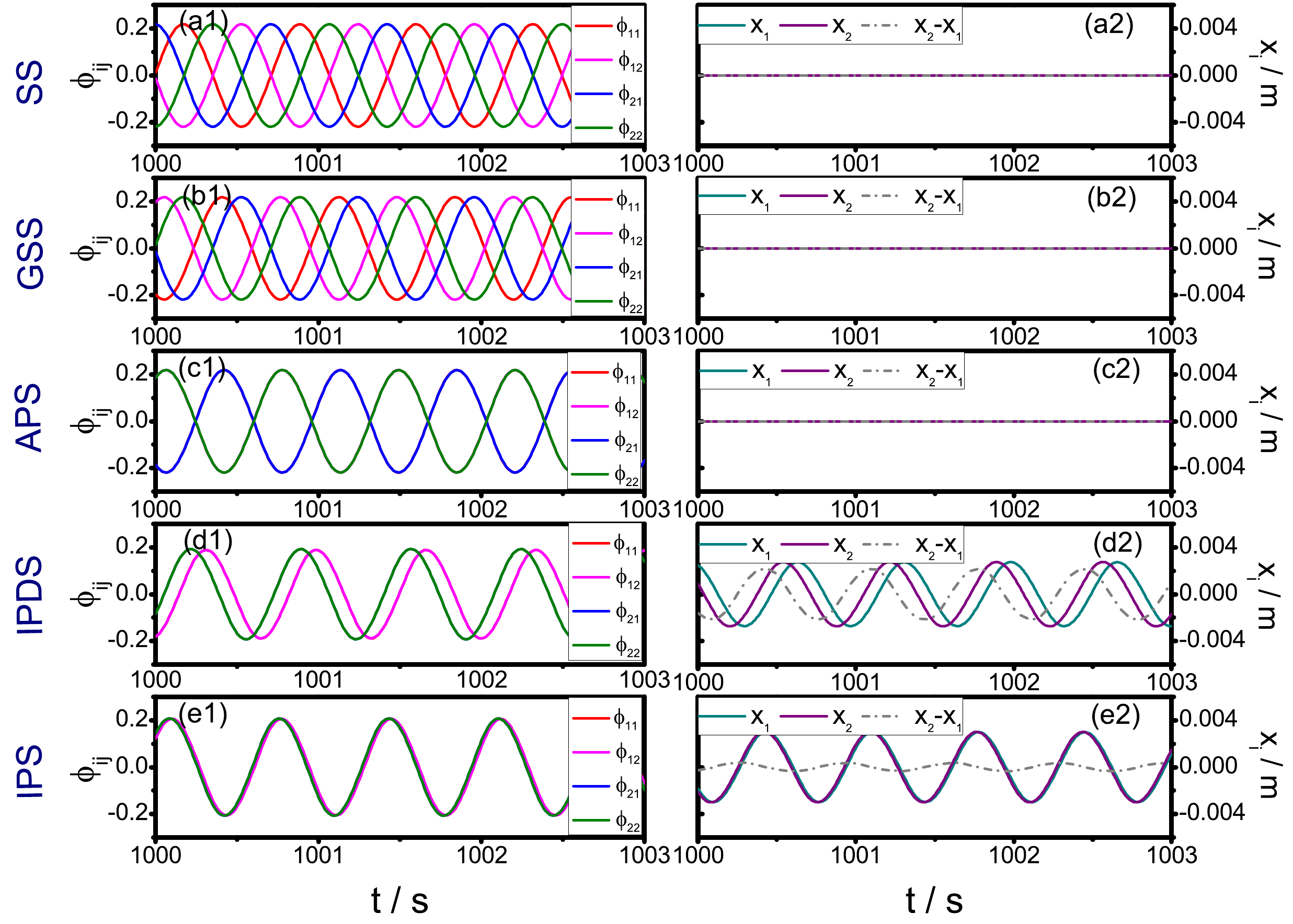}
\caption {(Color online) Synchronous patterns observed in numerical simulations. The initial conditions of the upper pendulums are fixed as $(\phi_{21},\phi_{22})=(-0.13,-0.27)$. By the frication coefficient $c_{x_{2}}=65$, (a) the splay synchronization (SS) state generated by the initial condition $(\phi_{11},\phi_{12})=(0.15,0.23)$, (b) the generalized splay synchronization (GSS) state generated by the initial condition $(\phi_{11},\phi_{12})=(-0.15,0.23)$, (c) the anti-phase synchronization (APS) state generated by the initial condition $(\phi_{11},\phi_{12})=(-0.15,0.05)$, and (d) the in-phase delay synchronization (IPDS) state generated by the initial condition $(\phi_{11},\phi_{12})=(-0.15,-0.23)$. (e) By $c_{x_{2}}=400$, the in-phase synchronization (IPS) state generated by the initial condition $(\phi_{11},\phi_{12})=(-0.15,-0.23)$. The left column shows the time evolution of the pendulum phases for each state. The right column show the movements of the supporting beams, as well as their displacement, for each state.}
\label{fig4}
\end{figure}

According to the experimental setup, in numerical simulations we treat the pendulums as identical, and remove the springs connected to the beams (so as to make the cardboard move freely). Specifically, we adopt the following set of parameters in numerical studies: $m_{ij}=m=1.0$ kg, $l_{ij}=l=0.124$ m, $c_{\phi_{ij}}=c=0.01$ N$\cdot$s$\cdot$m, $D=0.075$ N$\cdot$m, $M_{1}=20.0$ kg, $M_{2}=10.0$ kg, $c_{x_{1}}=5.0$ N$\cdot$s/m, $g=9.81$ m/s$^{2}$, $k_{x_{1}}=k_{x_{2}}=0$ N/m, and $\gamma_{N}=\pi/36$. The beams are initially staying at the equilibrium points with $0$ velocity, ${x}_{1}={x}_{2}=0$ m, ${\dot{x}}_{1}={\dot{x}}_{2}=0$ m/s. As in experiment, we set the pendulums with random initial phases and $0$ initial momentum. Eq. (3) is integrated by the fourth-order Runge-Kutta method with the time step $\delta t=1\times 10^{-3}$. We vary the initial conditions to search for all the synchronous patterns for the given set of parameters, and scan the phase space for the attracting basins of the patterns. The transitions of the synchronous patterns are investigated by changing the friction coefficient $c_{x_2}$.

Fixing $c_{x_2}=65$ N$\cdot$s/m, we plot in Figs. \ref{fig4}(a1-d1) all the synchronous patterns observed in simulation by varying the initial phases of the pendulums. Fig. \ref{fig4}(a1) plots the time evolution of the phases started from the initial condition $(\phi_{11},\phi_{12},\phi_{21},\phi_{21})=(0.15,0.23,-0.13,-0.27)$. It is seen that the phases are equally separated in the space by the phase difference $\Delta \phi=\pi/2$ and pendulums on each layer reach anti-phase synchronization, showing the feature of SS observed in experiment [Fig. \ref{fig2} (a)]. As anti-synchronization is established between pendulums on each layer, both beams are steady during the system evolution, $x_1=x_2=0$, as depicted in Fig. \ref{fig4}(a2). Fig. \ref{fig4}(b1) shows the time evolution of the phases started from a slightly different initial condition, $(\phi_{11},\phi_{12},\phi_{21},\phi_{21})=(-0.15,0.23,-0.13,-0.27)$. It is seen that the system is settled onto a state where pendulums on the same layer satisfy anti-phase synchronization, but pendulums on the different layers show a constant phase difference, $\Delta \phi\approx 0.51$. This is essentially the feature of GSS, as the one plotted in Fig. \ref{fig2}(b) based on experiment. Fig. \ref{fig4}(c1) is generated by the initial condition $(\phi_{11},\phi_{12},\phi_{21},\phi_{21})=(-0.15,0.05,-0.13,-0.27)$, where anti-phase synchronization is established between pendulums on the same layer, while pendulums on different layers may reach either in-phase or anti-phase synchronization. We thus identify this state as the APS state, which corresponds to the experimental results plotted in Fig. \ref{fig2}(c). As in both GSS and APS states metronomes on the same layer reach anti-phase synchronization, the two supporting beams thus are steady, as depicted in Figs. \ref{fig4}(b2,c2). Fig. \ref{fig4}(d1) shows the system dynamics started from the initial condition $(\phi_{11},\phi_{12},\phi_{21},\phi_{21})=(-0.15,-0.23,-0.13,-0.27)$. It is seen that the two pendulums on each layer reach in-phase synchronization, but their is a constant phase delay between pendulums on different layers. This state corresponds to the IPDS state observed in experiment [Fig. \ref{fig2}(d)]. IPDS is essentially different from the other states, as no anti-synchronization (which is believed to be a stable state in coupled metronomes) is observed among the pendulums. Fig. \ref{fig4}(d2) shows the movements of the two beams in IPDS state, where harmonic oscillations are observed. 

Motivated by the experimental studies, we increase the coupling strength between the two layers in order of generating IPS. In our model, this is realized by increasing the friction coefficient $c_{x_{2}}$. By the same initial conditions used in Fig. \ref{fig4}(d1), we plot in Fig. \ref{fig4}(e1) the system evolution by increasing $c_{x_{2}}$ to $400$ N$\cdot$s/m. It is seen that, during the course of system evolution, the phases are keeping almost identical ($\Delta \phi\approx 1\times 10^-2$), showing the character of IPS observed in experiment [Fig. \ref{fig2}(e)]. Further simulations also show that by increasing the friction coefficient, the small phase delay between the pendulums can be further decreased. This observation is reasonable, as with the increase of $c_{x_2}$ the binding of the two beams will be tightened, therefore the two beams can be regarded as one rigid beam for sufficiently large $c_{x_2}$. Fig. \ref{fig4}(e2) shows the movements of the two beams, as well as their displacement. Comparing to the IPDS state [Fig. \ref{fig4}(d2)], it is evident that the motions of the two beams are highly correlated. 

\begin{figure}
\includegraphics[width=\columnwidth]{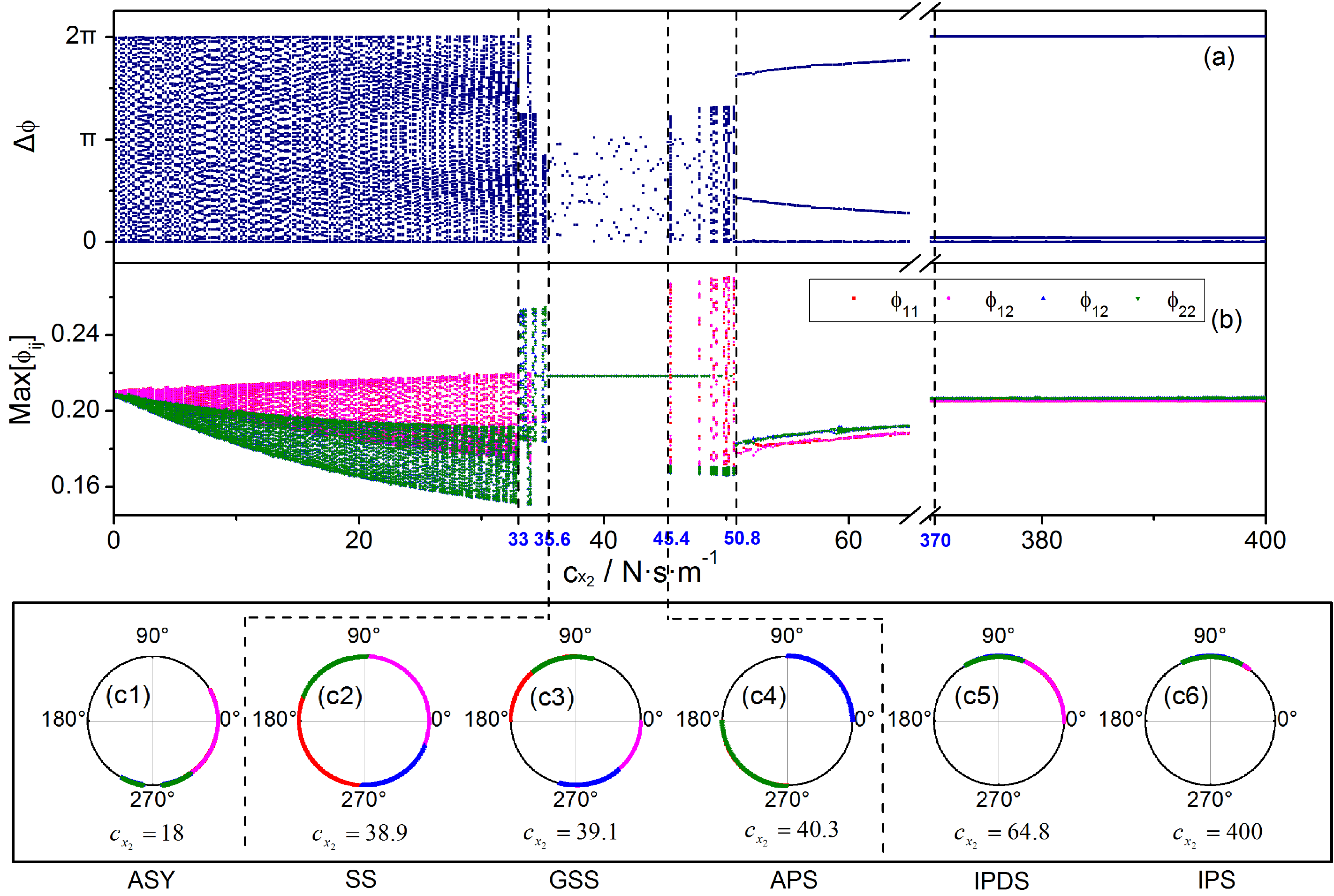}
\caption {(Color online) The transition of the system dynamics as a function of the friction coefficient $c_{x_{2}}$. The initial conditions are identical to that of Fig. \ref{fig4} (d) and (e). (a) The variation of the phase delay, $\Delta \phi$, between adjacent local phase maxima. (b) The variation of the phase amplitude, $\max{[\phi_{ij}]}$. Measure synchronization is observed at about $c_{x_2}=32$. Bottom panel: typical synchronous patterns observed in the transition process. In the bottom panel, each subplot is plotted by recording the trajectories of the pendulums for a short period. See context for more details on pattern transitions.} 
\label{fig5}
\end{figure} 

We go on to analyze the transition of the system dynamics as a function of the friction coefficient $c_{x_2}$. Fixing the initial conditions as $(\phi_{11},\phi_{12},\phi_{13},\phi_{14})=(-0.15,-0.23,-0.13,-0.27)$ [the same initial conditions used in Fig. \ref{fig4} (d) and (e)], we plot in Fig. \ref{fig5}(a) the variation of the phase delay, $\Delta \phi$, between adjacent ``ticks" (i.e., the local maximum of the phases) as a function of $c_{x_2}$. For each $c_{x_2}$, a period of $2$ seconds is monitored. For the asynchronous state, the phases are not clocked and $\Delta \phi$ will be varying chaotically with time. When SS is reached, $\Delta \phi$ will be fixed to $\pi/2$. In the GSS state, $\Delta \phi$ will be switching alternatively between two constant values, with the sum of them equals $\pi$. When APS is established, $\Delta \phi$ will be fixed to $\pi$. For IPDS, $\Delta \phi$ is switching alternatively between two constant values too. However, different from GSS, in IPDS the sum of the two values equals $2\pi$. Finally, in the IPS state we have $\Delta \phi \approx 0$. As shown in Fig. \ref{fig5}(a), with the increase of $c_{x_2}$, the distribution of $\Delta \phi$ is firstly continuous ($0<c_{x_2}<35.6$), then discrete ($35.6<c_{x_2}<47.8$), and, after a chaotic transition regime ($47.8<c_{x_2}<50.8$) shrunk to two branches (please note that $0$ is identical to $2\pi$ in phase). As $c_{x_2}$ increases further ($50.8<c_{x_2}<370$), the two branches are gradually combined into one branch. Based on the definition of $\Delta \phi$, Fig. \ref{fig5}(a) thus indicates the following scenario of pattern transition: asynchronous state $\rightarrow$  SS (GSS,APS) $\rightarrow$ asynchronous state $\rightarrow$ IPDS $\rightarrow$ IPS.

To explore the energy exchange among the pendulums \cite{2012 Kapitaniak}, we monitor the variation of the phase amplitudes, $\max{[\phi_{ij}]}$, as a function of the friction coefficient $c_{x_2}$. The results are plotted in Fig. \ref{fig5}(b). It is seen that before the system is transited from the asynchronous to synchronous states (around $c_{x_2}=35.6$), the phase amplitudes are changed from non-identical to identical (around $c_{x_2}=32$), indicating the occurrence of measure synchronization around this point (the time-averaged energy of the pendulums are identical) \cite{MS}. For this observation, we conjecture that measure synchronization plays as an indicator for the emergence of synchronous patterns. Measure synchronization is maintained till the transition regime is reached ($47.8<c_{x_2}<50.8$). After that, measure synchronization is gradually resumed as $c_{x_2}$ increases. To have more details on the transition of the system dynamics with respect to the friction coefficient, we plot in the bottom panel of Fig. \ref{fig5} the typical states observed during the transition process, which demonstrate vividly how the system dynamics is transited among different synchronous patterns as the friction coefficient varies.

\begin{figure}
\center
\includegraphics[width=\columnwidth]{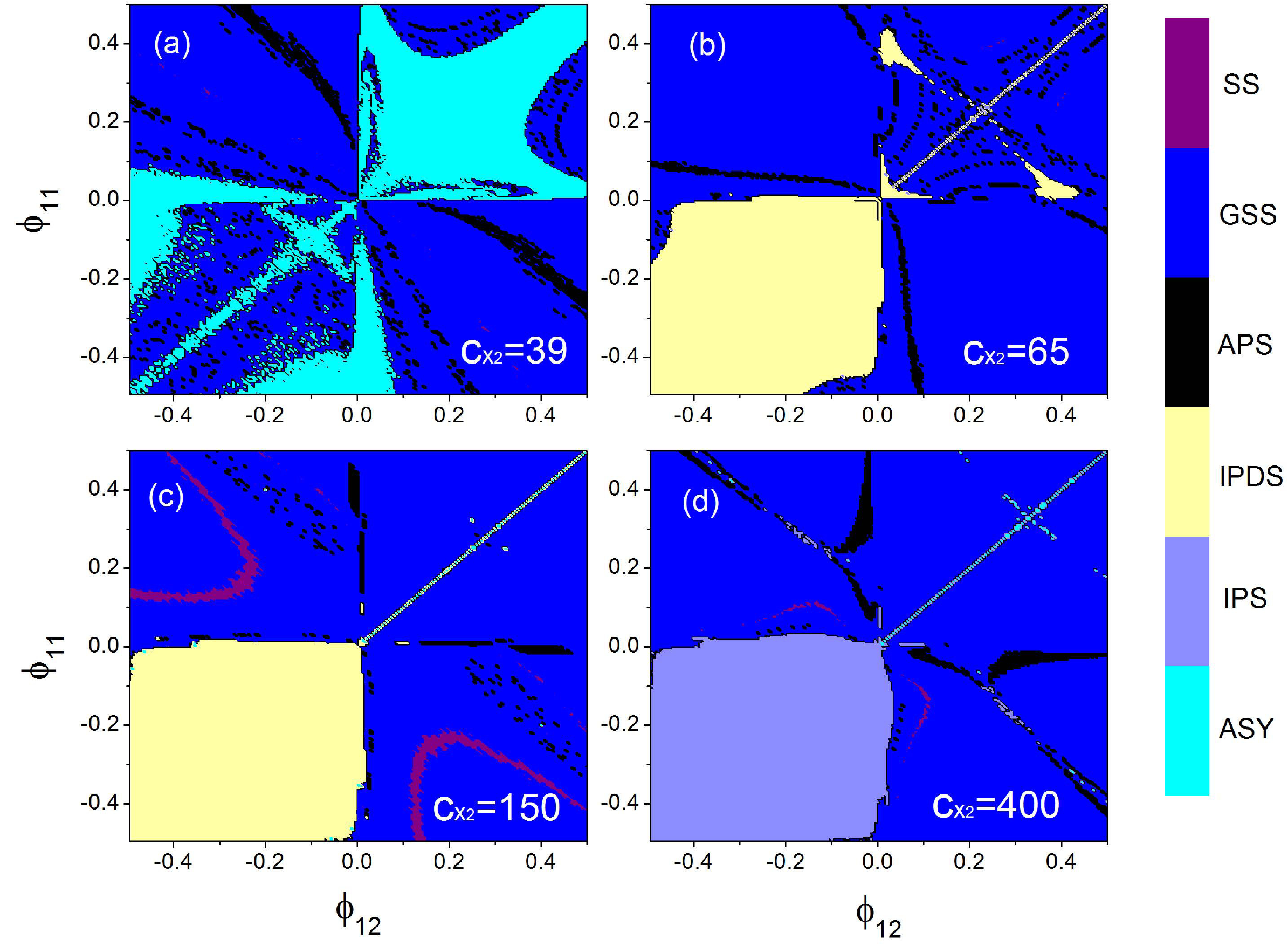}
\caption {(Color online) The attracting basins of the synchronization states in the space $(\phi_{11},\phi_{12})$ for different friction coefficients, $c_{x_2}$. The initial conditions of the upper pendulums are fixed as $(\phi_{21},\phi_{22})=(-0.13,-0.27)$. (a) $c_{x_2}=39$. (b) $c_{x_2}=65$. (c) $c_{x_2}=150$. (d) $c_{x_2}=400$. The system states are colored as follows: splay synchronization state (SS, purple), generalized splay synchronization state (GSS, light blue), anti-phase synchronization state (APS, blue), in-phase delay synchronization state (IPDS, light yellow), in-phase synchronization state (IPS, light purple), and asynchronous state (ASY, cyan).}
\label{fig6}
\end{figure}

Motivated by the fact that different patterns are generated with different probabilities, we finally investigate the attracting basins of the synchronous patterns. To reduce the dimension of the scanning space, in our studies we fix $(\phi_{21},\phi_{22})=(-0.13,-0.27)$, while changing $\phi_{11}$ and $\phi_{12}$ within the interval $[-0.5,0.5]$ to explore the basins. Fixing $c_{x_2}=39$, we plot in Fig. \ref{fig6}(a) the distribution of the basins, with each system state (asynchronous state, SS, GSS, APS, IPDS, IPS) represented by an individual color. It is seen that the space is dominated by the asynchronous and GSS states, scattered with small islands of APS state. Please note that due to the symmetry of the upper pendulums (the exchange of them does not change the system dynamics), the basin distribution is symmetric along the axis $\phi_{21}=\phi_{22}$. Increasing $c_{x_2}$ to $65$, we scan the attracting basins again, and the results are plotted in Fig. \ref{fig6}(b). Comparing to Fig. \ref{fig6}(a), it is seen that the asynchronous state is disappeared, and the IPDS and GSS states are dominant in the space. Fig. \ref{fig6}(c) shows the result for $c_{x_2}=150$. Here it is seen that SS is newly appeared, which, like APS, occupies only a small volume (small attracting basin) in the space. Increasing $c_{x_2}$ further to $400$, Fig. \ref{fig6}(d) shows that the IPDS state, which is dominant in Fig. \ref{fig6}(c), is disappeared and replaced by the IPS state. These numerical findings confirm the experimental observation that the distribution of the synchronous patterns is crucially dependent on the coupling strength of the two layers (the larger is the friction coefficient $c_{x_2}$, the stronger is the coupling strength). Fig. \ref{fig6} also reveals the role of GSS played in pattern transitions: it bridges the different patterns, i.e., the basins of other synchronous patterns are separated from each other by GSS states. The results shown in Fig. \ref{fig6} deepen our understanding on the transition of the system dynamics in two aspects: (1) As the attracting basins are distributed irregularly in the phase space, the transition path therefore will be dependent on the initial conditions. This finding is consistent with the numerical observations [Fig. \ref{fig5}]. (2) The transition between different types of synchronization patterns is generally mediated by GSS. Taken the situation of Fig. \ref{fig6}(c) as an example, under noise perturbations, the system may change from SS to APS via GSS. (A detailed analysis on the role of GSS on synchronization transition, as well as the stability of the synchronous patterns, will be reported elsewhere).

\section{Discussions and conclusion}\label{s.6}

The asymmetric coupling scheme offers a new approach for exploring the collective behaviors of coupled metronomes. While coupling scheme has been proven to be an important element for oscillator synchronization, its role on metronome synchronization has not be explored. As shown in the present study, the adoption of asymmetric coupling is able to generate new patterns which are absent in the symmetric coupling scheme. It is reasonable to expect that by breaking the coupling symmetry in new manners, more interesting synchronization behaviors would be observed. Here, by distributing four metronomes over two different layers, the system symmetry is deteriorated from global to local symmetry. Along this line, we may generate more synchronization patterns by introducing new asymmetries, e.g., introducing parameter mismatch among the metronomes \cite{ClockSyn:nonidentical}, configuring the coupling strength in an asymmetric fashion \cite{,AsymmCoupling}, using biased couplings \cite{BiasedCoupling}, etc. Meanwhile, in the present study, for the purpose of demonstration, we have employed only four metronomes and two layers, richer phenomena are expected by increasing the number of metronomes and layers. 

We wish to highlight the audio-based technique used in detecting the synchronous patterns. Comparing to the traditional sophisticated video-based techniques, this new technique is more convenient and operational. Moreover, as the metronomes tick only at their extremes, the audio signals are featured by distinctly separated pulses, which are much larger than the noise background. This feature makes the identification of the synchronous pattern more accurate and reliable. The small trick we have used in differentiating the audio signals of the upper metronomes from the lower ones, i.e., covering the speakers of the upper metronomes by cotton cloth, is very helpful in detecting the configuration of the synchronous patterns. Without this treatment, we can not differentiate the signals of the upper metronomes from the lower ones. We hope these techniques, probably after some improvements, could be used in future studies for exploring the more complicated behaviors emerged in coupled metronomes.  

To summarize, we have studied, theoretically and experimentally, the synchronization behaviors of coupled metronomes distributed on two layers. By varying the initial conditions of the metronomes and the coupling strength between the two layers, we have observed rich synchronous patterns, as well as the transition between the patterns. In particular, a new type of pattern, namely the IPDS state, is observed for the first time, which is rooted in the breaking of the coupling symmetry in this layered coupling scheme. By numerical simulations, we have analyzed the basin distribution of the patterns, and investigated in detail the variation of the basins as a function of the coupling strength between the two layers, from which a global picture on the pattern transition is portrayed. Additionally, a new measure technique based on audio signals has been proposed for detecting the synchronous patterns in experiment, which, comparing to the traditional video-based techniques, is more convenient and operational. Our study highlights the significant impact of coupling symmetry on the formation of synchronous patterns, and the new experimental setup could be used in classroom for demonstration purposes.

This work was supported by the National Natural Science Foundation of China under Grant No. 11375109, and also by the Fundamental Research Funds for the Central Universities under Grant No. GK201601001.

\end{document}